\documentstyle[11pt,newpasp,twoside,epsf]{article}
\begin{document}

\title{The Properties of Spiral Galaxies in 
	Semi-Analytic Galaxy Formation Models}
\author{Eric F.~Bell}
\affil{Steward Observatory, 933 N. Cherry Ave., Tucson, AZ 85721, U.S.A.}
\author{Carlton M. Baugh, Shaun Cole, Carlos S. Frenk and Cedric G. Lacey\altaffilmark{1}}
\affil{University of Durham, Dept. of Physics, South Road, Durham, 
	DH1 3LE, U.K.}
\altaffiltext{1}{SISSA, via Beirut 2-4, 34014 Trieste, Italy}

\begin{abstract}
We examine the present-day galaxy disk colors and star
formation rates (SFRs) in the semi-analytic model of 
Cole et al. (2000).  We find that the fiducial model
is a good match to Kennicutt's (1998) observed global 
star formation law, the color-based metallicity--magnitude
correlation and the colors of low-luminosity galaxies.
The main limitation of the model, from the point of view
of present-day spirals, is that the optical colors of the disks of 
very luminous spiral galaxies are too blue, even after accounting 
for the effects of dust.
\end{abstract}

\begin{figure}
\begin{center}
\begin{picture}(300,150)
\put(-40,0)
{\epsfxsize=6.0truecm \epsfysize=6.0truecm 
\epsfbox[200 250 425 485]{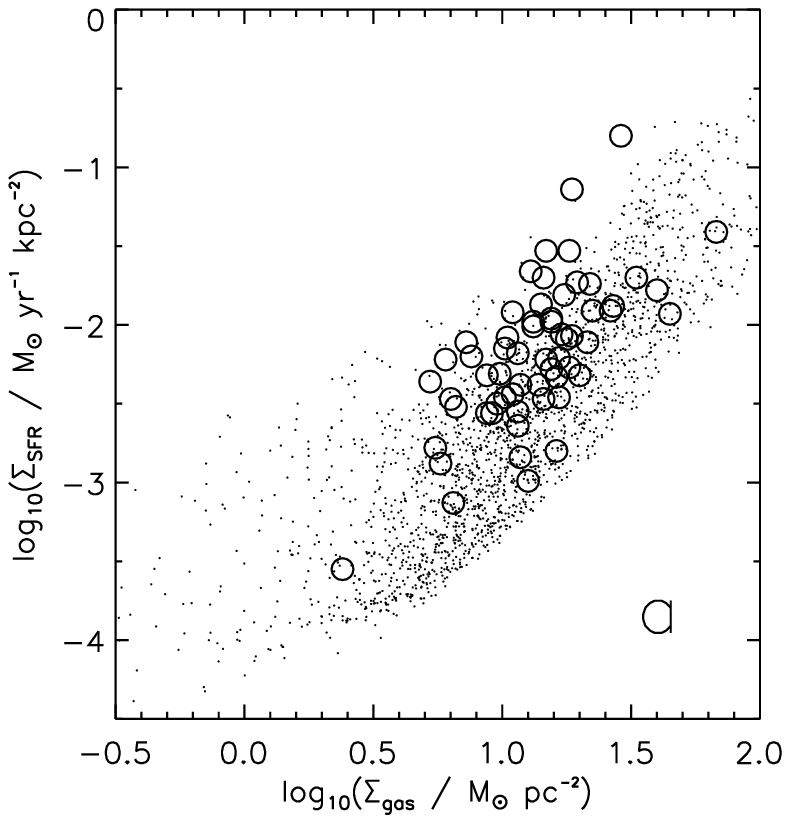}}
\put(140,0) 
{
\epsfxsize=6.0truecm \epsfysize=6.0truecm 
\epsfbox[200 250 425 485]{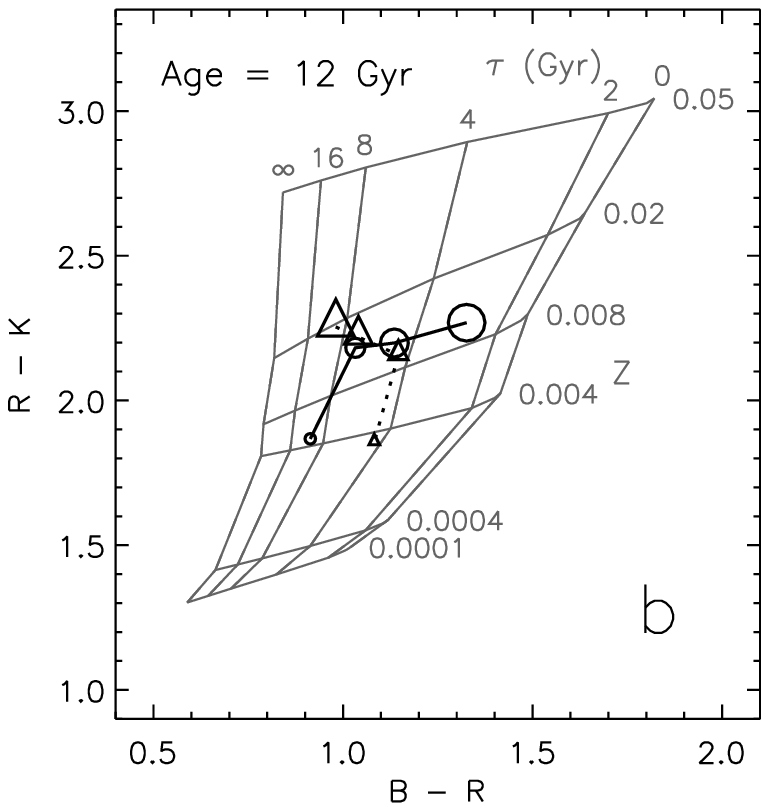} 
}
\end{picture}
\caption{
{\bf a)} The
SFR per unit area against the gas density per unit area for the 
observed data (circles) and model galaxies (dots).  
{\bf b)} Comparing the median optical--near-IR colors of 
the models (open triangles connected by the dotted line) and 
observations (open circles connected by the solid line).  
Symbols are coded by magnitude: larger symbols
for bright galaxies and smaller symbols for faint galaxies.
Overplotted is a stellar population model grid for various $\tau$ models
($\tau = 0$ is a single burst; $\tau = \infty$ has a constant SFR)
with a range of metallicities ($Z = 0.02$ is solar; 
Bruzual \& Charlot, in preparation).
}
\end{center}
\end{figure}

Semi-analytic galaxy formation is a powerful technique 
with which to explore the most important physical processes in 
galaxy formation and evolution (such as dark matter
halo formation and evolution, gas cooling, star formation, the
return of chemical elements and energy back into the interstellar
medium, and dust obscuration).  
One of the most important assumptions
underlying these models is that all galaxies either have been or
presently are disk-dominated.  Consequently, the properties of spiral
galaxies in this type of model are of special importance.  
Here, we compare the colors (accounting for dust) 
and SFRs of present-day spiral galaxies in the model of Cole et al. (2000) 
with observational constraints: Cole et al. (2000) have  
already demonstrated that the luminosities and sizes of 
their model disk galaxies are in good agreement with observations.

We compare the colors of the model disks
with those of a sample of 121 low-inclination 
spiral galaxies taken from Bell \& de Jong (2000).  We estimate
colors at one disk half-light radius: these colors should be 
representative of the whole disk and should be only minimally
affected by dust.  Observational 
data for the SFR comparison was taken from Kennicutt (1998).

In panel a of Fig.~1 we show the comparison 
of the observed and model galaxy SFRs.  Model SFRs
are scaled by the disk size, accounting for the differing
distribution of gas, young stars and old stars in observed spiral galaxies.
The model SFRs provide a surprisingly good match to the observed galaxies:
the SFR in the model does not depend on the gas density
(only on the global dynamical timescale, 
the cold gas mass and disk circular velocity).  There is 
a slight offset between the model and observed SFRs: this may be
removed when model galaxies are selected in the same way as 
the observational sample, or by using a slightly different 
spatial distribution of gas and the old and young stars.

In panel b we show the median colors of 
observed and model galaxies as a function of magnitude (large symbols
$M_K < -24.5$; small symbols $M_K > -21.5$).
Model galaxies are selected to have the same $K$-band surface brightnesses
and absolute magnitudes as the observed galaxies, ensuring that 
we are comparing model disks to their observational analogues.
The $R - K$ colors, which are primarily
a metallicity indicator, are accurately reproduced by the model galaxies.
Also, the $B - R$ colors of model low-luminosity
spiral disks are a reasonable match to the observations.  However, 
the $B - R$ colors of the most luminous model disks are too blue,
indicating that the model disks have a young luminosity-weighted
age.   We are working towards understanding this discrepancy:
two possibile resolutions include modifying the star formation law and 
altering the recipe for gas cooling at late times in the largest 
galaxy-sized halos.

\acknowledgements

Support for EFB was provided by NASA LTSA grant NAG5-8426 and
NSF grant AST-9900789.

\end{document}